%% file: 1QBit_Prime_Factorization_Research_Paper.tex
\documentclass[12pt,threecolumn]{article}

\usepackage{amssymb, amsmath, amscd}
\usepackage{graphicx}
\usepackage[latin1]{inputenc}
\usepackage{epsfig}
\usepackage{color}
\usepackage{rotating}

\numberwithin{equation}{section}
\usepackage[affil-it]{authblk}

\begin{document}
\input{macro}

\bibliographystyle{amsalpha}

\title{Prime factorization using quantum annealing and computational algebraic geometry}

\author{Raouf Dridi%
  \thanks{\texttt{raouf.dridi@1qbit.com}} \
  and  
  Hedayat Alghassi%
  \thanks{\texttt{hedayat.alghassi@1qbit.com}} }
\affil{1QB Information Technologies (1QBit)\\
       458--550 Burrard Street\\
       Vancouver, British Columbia, Canada \, V6C 2B5}

  \date{\today}
\maketitle

\begin{abstract}
	~~\\ 
	 We investigate prime factorization from 
	 two perspectives: quantum annealing and computational algebraic geometry, specifically   Gr\"obner bases. 
	 We present a novel autonomous algorithm which combines the two approaches and leads
	 to the factorization of all bi-primes up to just over \mbox{200 000}, the largest number factored to date using a quantum processor.  
	 We also explain how  Gr\"obner bases can be used to reduce
	 the degree of Hamiltonians.  
\end{abstract}

 \section{Introduction}
 
Prime factorization is at the heart of secure data transmission because it is widely believed to be NP-complete. In the prime factorization problem, for a large bi-prime $M$, the task is to find the two prime factors $p$ and~$q$ such that $M=pq$. 
In secure data transmission, the message to be transmitted is encrypted using a public key which is, essentially,  a large bi-prime that can only be decrypted using its prime factors, which are kept in a private key. 
Prime factorization also connects  to many branches of mathematics; 
two branches relevant to us are computational algebraic geometry and quantum annealing.  

~~\\ 
To leverage the problem of finding primes $p$ and $q$ into the realm of computational algebraic geometry, it suffices  to transform it into an algebraic system of equations $\mathcal S$. This is done using the binary representation $p = 1+\sum_{i=1..s_p} 2^i P_i \mbox{ and }q=1+\sum_{i=1..s_q} 2^i Q_i$, which is plugged into $M=pq$ and expanded into a system of  polynomial equations. The reader is invited to read the sections \emph{Methods 4.1} and \emph{4.2} for the details of this construction. The system $\mathcal S$ is  given by this initial system of equations in addition to  the auxiliary equations expressing the binary nature of the variables $P_i$ and $Q_i$, carry-on, and connective variables. The two primes $p$ and $q$ are then   given by the unique zero of~$\mathcal S$.
In theory, we can solve the system~$\mathcal S$ using Gr\"obner bases; however, in practice, this alone does not work, since Gr\"obner basis computation~(Buchberger's algorithm) is exponential.
 
 ~~\\  
The connection to quantum annealing can also be easily described. Indeed, finding $p$ and~$q$ can be formulated into an unconstrained binary optimization problem~$(\mathcal P)$,  where the cost function $f$ is the sum of the squares of polynomials in $\mathcal S$.  The unique zero of~$\mathcal S$ now sits on the unique global minimum of $(\mathcal P)$ (which has  minimum energy equal to zero). There are, however, a few non-trivial requirements we need to deal with before solving the cost function using quantum annealing. These
 requirements  concern the nature of cost functions that quantum annealers can handle. In particular, we would like the cost function of $(\mathcal P)$ to be a positive quadratic polynomial. We also require that the coefficients of the cost function 
be rather uniform and match the hardware-imposed dynamic range.  

~~\\ 
In the present paper, we suggest looking into the problem through both lenses, and demonstrate that indeed this
 approach gives better results. In our scheme, we will be using quantum annealing to solve~$(\mathcal P)$, but at the same time  we will be using  
Gr\"obner bases to help us reduce the cost function~$f$  into  a positive quadratic polynomial $f^+$ with desired values for the coefficients.  
We will be also using  Gr\"obner bases at the important step of pre-processing $f^+$ before finally passing it to the quantum annealer. This pre-processing significantly reduces the size of the problem. 
The result of this
combined approach is an algorithm with which we have been able to factorize all bi-primes up to~$2\times 10^5$ using the D-Wave 2X processor. The algorithm is autonomous in the sense that no a priori knowledge, or manual or ad hoc pre-processing, is involved. We refer the interested reader to 
{\it Supplementary materials 5} for a brief description of the D-Wave 2X processor, along with some statistics  for several of the highest numbers that we embedded and solved.  
More detail about the processor architecture can be found in 
\cite{natueDwave}.  Another important reference is the work of S. Boixo et al. in \cite{BoixoNature}, which  presents  experimental evidence that the scalable D-Wave  processor implements quantum annealing with surprising robustness against noise and imperfections.  Evidence that, during a critical portion of
quantum annealing, the qubits become entangled and entanglement persists even as the system reaches equilibrium is presented in \cite{PhysRevX.4.021041}.

 ~~\\ 
Relevant to us also is the work in \cite{realalggeo}, which uses algebraic geometry to solve optimization problems (though not specifically factorization;  see \emph{Methods 4.4} for an adaptation to factorization). Therein,  Gr\"obner bases are  used to compute standard monomials and transform the given optimization problem into an eigenvalue computation. 
Gr\"obner basis computation is the main step in this approach, which makes it inefficient, given  the high complexity of Gr\"obner basis computation. In contrast to that work, we ultimately solve the optimization problem using a quantum annealing processor and pre-process and adjust the problem with algebraic tools, that is, we reduce the size of the cost function and adjust the range of its parameters. However, we share that work's point of view of using real algebraic geometry, and our work is the first to introduce algebraic geometry, and  Gr\"obner bases in particular, to solve quantum annealing-related problems. We think that this is a fertile direction for both practical and theoretical endeavours. 
 
~~\\ 
Mapping the factorization problem into a degree-4 unconstrained binary optimization problem is first discussed
in \cite{microsoft}. There, the author proposes  solving the problem using a continuous optimization method he calls curvature inversion descent. Another related work is the quantum annealing factorization algorithm proposed in  \cite{DBLP:journals/qic/SchallerS10}. We will discuss it in the next section and improve upon it in two ways. The first involves the addition of the pre-processing stage using Gr\"obner bases of the cost function. This dramatically reduces the number of variables therein. The second way concerns the reduction of the initial cost function, for which we propose a general Gr\"obner basis scheme that precisely answers the various requirements of the cost function. In \emph{Results 2.2}, we present our algorithm (the column algorithm) which outperforms this improved algorithm (i.e., the cell algorithm).  Using a reduction proposed in  \cite{DBLP:journals/qic/SchallerS10} and ad-hoc simplifications and tricks, the paper \cite{crapyPRL}  reports the factorization of bi-prime 143 on a liquid-crystal NMR quantum processor,  which until now was the largest reported bi-prime number factored in any quantum computation realization.  

~~\\ 
This review is far from complete without mentioning Shor's  algorithm \cite{MR1471990} and Kitaev's phase estimation \cite{hsp}, which, respectively, solve the factorization problem and the abelian hidden subgroup problem in polynomial time, both for the gate model paradigm. The largest number factored using a physical realization of Shor's algorithm is 15 \cite{Monz1068}; see \cite{smolin} also for a discussion about oversimplification in the previous  realizations. 
Finally, in \cite{PhysRevA.88.022322}, it has been proved that contextuality is needed for any speed-up in a measurement-based quantum computation factorization algorithm.

\section{Results}
The binary multiplication of the two primes $p$ and $q$ can be expanded in two ways: cell-based and column-based procedures (see \emph{Methods 4.1} and \emph{4.2}). 
Each procedure leads to a different unconstrained binary optimization problem.  The cell-based procedure creates the unconstrained binary quadratic programming problem 
\begin{system}
[(\mathcal P_1)]  min_{ \mathbb Z_2}\, \sum_{ij} {H_{ij}} ^2,\\[3mm]
\mbox{with } H_{ij} := Q_{{i}}P_{{j}}+S_{{i,j}}+Z_{{i,j}}-S_{{i+1,j-1}}-2\,Z_{{i,j+1}},
\end{system}%
and the column-based procedure results in the problem	
\begin{system}
[(\mathcal P_2)]  min_{\mathbb Z_2}\,
	 \sum_{1\leq i\leq (s_p+s_q+1)} {H_{i}} ^2,\\[3mm]
\mbox{with } H_{i} := \sum\limits_{j = 0}^{sq} {{Q_j}} {P_{i - j}} + \sum\limits_{j = 1}^i {{Z_{j,i}} - {m_i}}  - \sum\limits_{j = 1}^{s_q +1 + i - m_i} {{2^{j - i}}{Z_{i, i+j}}}.
\end{system}%
The two problems $(\mathcal P_1)$ and $(\mathcal P_2)$ are equivalent.  Their  cost functions  
are not in quadratic form, and thus must be reduced before being solved using a quantum annealer.  
The reduction procedure is not a trivial task. 
In this paper we define,  for both scenarios: 1)  a reduced quadratic positive cost function and 2) a pre-processing procedure.  Thus, we present two different quantum annealing-based prime factorization algorithms. The first algorithm's decomposition method (i.e., the cell procedure, \emph{Methods 4.1}) has been addressed in  \cite{DBLP:journals/qic/SchallerS10}, without pre-processing and without the use of  Gr\"obner bases in the reduction step.  Here, we discuss it from  the Gr\"obner bases framework and add the important step of pre-processing. The second algorithm, however, is novel in transformation of its quartic terms to quadratic, outperforming the first algorithm due to its having fewer variables.  

~~\\ 
We write $\mathbb R[x_1, \ldots, x_n]$ for the ring of polynomials
in $x_1, \ldots, x_n$ with real coefficients and $\mathcal V(f)$ for the affine variety defined by 
the polynomial $f\in \mathbb R[x_1, \ldots, x_n]$, that is, the set of zeros of the equation $f=0$. Since we are interested only in the binary zeros (i.e., $x_i\in \mathbb Z_2$), we need
to add the binarization polynomials $x_i(x_i-1)$, where $i=1,...,n,$ to~$f$ and obtain the system $\mathcal S=\left\{f,\,  x_i(x_i-1), \, i=1,...,n\right\}$. 
The  system  $\mathcal S$ generates an ideal $\mathcal I $ by  taking all linear combinations over $\mathbb R[x_1, \ldots, x_n]$ of all polynomials in  $\mathcal S$; we have $\mathcal V(\mathcal S)=\mathcal V(\mathcal I).$ The ideal $\mathcal I$ reveals the hidden polynomials which are the consequence of the generating  polynomials in $\mathcal S$. To be precise, the set of all hidden polynomials is given
by the so-called radical ideal $\sqrt{\mathcal I}$, which is defined by \mbox{$\sqrt{\mathcal I} = \{g \in\mathbb R[x_1, \ldots, x_n] |\,  \exists r\in \mathbb N: \, g^r\in \mathcal I \}$}.  In practice,  the ideal $\sqrt{\mathcal I}$ is infinite, so we represent such an ideal
 using a Gr\"obner basis $\mathcal B$ which one might take to be a triangularization of the ideal $\sqrt{\mathcal I}$.  In fact, the computation of   Gr\"obner bases  generalizes Gaussian elimination in linear systems. We also have $\mathcal V(\mathcal S)=\mathcal V(\mathcal I)=\mathcal V(\sqrt{\mathcal I})=\mathcal V(\mathcal B)$  and $\mathcal I(\mathcal V(\mathcal I))=\sqrt{\mathcal I}$. 
 A brief review of Gr\"obner bases is given in \emph{Methods 4.3}.

\subsection{The cell algorithm}\label{cellalgo}
Suppose we would like to define 
the variety $\mathcal V(\mathcal I)$
by the set of  global minima of an unconstrained optimization problem 
$min_{\mathbb Z_2^n}(f^+)$, where $f^+$ is a quadratic polynomial. For instance, we would like $f^+$ to behave like $f^2$.  
Ideally, we want~$f^+$ to remain in $\mathbb R[x_1, \ldots, x_n]$ (i.e., not in a larger ring), which implies that no slack variables will be added. We also want~$f^+$ to
satisfy the following requirements:  
\begin{itemize}
    \item[(i)] $f^+$ vanishes on $V(\mathcal I)$ or, equivalently, $f^+\in\sqrt{\mathcal I}$.   
         
	\item[(ii)] $f^+ >0$ outside $V(\mathcal I)$, that is,  $f^+>0$ over $\mathbb Z_2^n-V(\mathcal I)$. 
	\item[(iii)] Coefficients of the polynomial $f^+$ are adjusted with respect to the dynamic range allowed by the quantum processor. 
\end{itemize}
Let $\mathcal B$  be a Gr\"obner basis for $\mathcal I$. We can then go ahead and define 
$$
	f^+ = \sum_{t\in \mathcal B|\, deg(t)\leq 2} a_t t,
$$
where the real coefficients $a_i$ are  subject to the requirements above; note that we already have~$f^+ \in \sqrt{\mathcal I}$ and thus the first requirement (i) is satisfied.  

~~\\
Let us apply this procedure
to  the optimization problem $(\mathcal P_1)$ above.  There, $f=H_{ij}$ and the ring of polynomials is $
	\mathbb R[P_{{j}},Q_{{i}},S_{{i,j}},S_{{i+1,j-1}},Z_{{i,j}},Z_{{i,j+1}}]. 
$
 We obtain the following Gr\"obner basis:
 
\begin{system}
 t_1 &:= &Q_{{i}}P_{{j}}+S_{{i,j}}+Z_{{i,j}}-S_{{i+1,j-1}}-2\,Z_{{i,j+1}}\label{HijInB},\\[2mm]
 t_2 &:= &\left( -Z_{{i,j+1}}+Z_{{i,j}} \right) S_{{i+1,j-1}}+ \left( Z_{{i,j+1
}}-1 \right) Z_{{i,j}},\\[2mm]
 t_3 &:= & \left( -Z_{{i,j+1}}+Z_{{i,j}} \right) S_{{i,j}}+Z_{{i,j+1}}-Z_{{i,j+1
}}Z_{{i,j}},\\[2mm]
 t_4 &:= &\left( S_{{i+1,j-1}}+Z_{{i,j+1}}-1 \right) S_{{i,j}}-S_{{i+1,j-1}}Z_{
{i,j+1}},\\[2mm]
 t_5 &:= &\left( -S_{{i+1,j-1}}-2\,Z_{{i,j+1}}+Z_{{i,j}}+S_{{i,j}} \right) Q_{{
i}}-S_{{i,j}}-Z_{{i,j}}+S_{{i+1,j-1}}+2\,Z_{{i,j+1}},\\[2mm]
 t_6 &:= &\left( -S_{{i+1,j-1}}-2\,Z_{{i,j+1}}+Z_{{i,j}}+S_{{i,j}} \right) P_{{
j}}-S_{{i,j}}-Z_{{i,j}}+S_{{i+1,j-1}}+2\,Z_{{i,j+1}},\\[2mm]
 t_7 &:= &\left( -Z_{{i,j+1}}+Z_{{i,j+1}}Z_{{i,j}} \right) Q_{{i}}+Z_{{i,j+1}}-
Z_{{i,j+1}}Z_{{i,j}},\\[2mm]
 t_8 &:= &-S_{{i+1,j-1}}Z_{{i,j+1}}+S_{{i+1,j-1}}Q_{{i}}Z_{{i,j+1}},\\[2mm]
 t_9 &:= &\left( -Z_{{i,j+1}}+Z_{{i,j+1}}Z_{{i,j}} \right) P_{{j}}+Z_{{i,j+1}}-
Z_{{i,j+1}}Z_{{i,j}},\\[2mm]
 t_{10} &:= &-S_{{i+1,j-1}}Z_{{i,j+1}}+S_{{i+1,j-1}}P_{{j}}Z_{{i,j+1}}.
\end{system}%
We have used the lexicographic order 
 $plex(P_{{j}},Q_{{i}},S_{{i,j}},S_{{i+1,j-1}},Z_{{i,j}},Z_{{i,j+1}})$; see {\it Methods 4.3} for definitions. 
Note that $t_1 = H_{ij}$.  We   define 
$$
	H_{ij}^{+} = \sum_{t\in \mathcal B|\, deg(t)\leq 2} a_t t, \mbox{ that is, } H_{ij}^{+}=\sum_{1\leq i\leq 6} a_i t_i,
$$
where the real coefficients $a_i$ are to be found. We need to constrain the coefficients
$a_i$ with the other requirements.  
The second requirement (ii), which translates into a set of inequalities on the unknown
coefficients $a_i$, can be obtained through a brute force evaluation of $H_{ij}$ and $H_{ij}^{+}$ over  the $2^6$ points of $\mathbb Z_2^6$. The outcome of this evaluation is a set of inequalities expressing the second requirement (ii)  (\emph{see Supplementary materials 5}). 

~~\\ 
The last requirement (iii) can be expressed in different ways. We can, for instance, require
that the absolute values of the coefficients of  ${H_{ij}^{+}}$,  with respect to the variables  $P_{{j}},Q_{{i}},S_{{i,j}},S_{{i+1,j-1}},Z_{{i,j}}$, and $Z_{{i,j+1}}$,
be within $[1-\epsilon, \, 1+\epsilon]$. This, together with the set of inequalities from the second requirement, define a continuous optimization problem and can be easily solved.  Another option is to minimize the distance between the coefficients to one coefficient. 
The different choices of the objective function and the solution of the corresponding continuous optimization problem are presented in \mbox{\emph{Supplementary materials 5}}.   

~~\\ 
Having determined the quadratic polynomial $H_{ij}^{+}\in R$ satisfyies the important requirements (i, ii, and iii) above, we 
can now phrase our problem $(\mathcal P_1)$ as the equivalent quadratic unconstrained binary optimization problem 
$
	min_{ \mathbb Z_2}\, \sum_{ij} {H_{ij}^{+}}. 
$
Notice that this reduction is performed only once for all cases; it need not to be redone for different bi-primes $M$.  
Before passing the problem to the quantum annealer, we use   Gr\"obner bases to reduce the size of the problem. In fact, what we pass to the quantum annealer is  
$
	\mathcal H = \sum  {\sf NF}_\mathcal B \left (H_{ij}^{+} \right)
$,
where {\sf NF} is the normal form and $\mathcal B$ is now the Gr\"obner basis cutoff, which we discuss in the next section. 
The largest bi-prime number that we embedded and solved successfully using the cell algorithm is $\sim${35~000}.
The following table presents some bi-prime numbers $M$ that we factored using the cell algorithm,  the number of variables using both the customized reduction {\it CustR} and the window-based {\it GB} reduction, the overall reduction percentage {\it R\%}, and the embedding and solving status inside the D-Wave 2X processor {\it Embed}. 

\[\mathop {\boxed{\begin{array}{*{20}{c}}
  {\underline M }&{\underline {p \times q} }&{\underline {CustR} }&{\underline {GB} }&{\underline {R \textit{\%}} }&{\underline {Embed} } \\ 
  {31861}&{211 \times 151}&{111}&{95}&{14}&\surd  \\ 
  {34889}&{251 \times 139}&{111}&{95}&{14}&\surd  \\ 
  {46961}&{311 \times 151}&{125}&{109}&{13}& \times  \\ 
  {150419}&{431 \times 349}&{143}&{125}&{12}& \times  
\end{array}}}\limits^{\begin{array}{*{20}{c}}
  {{\mathbf{Cell}}}&{{\mathbf{algorithm}}} 
\end{array}} \]


\subsection{The column algorithm  (factoring up to 200 000)}\label{betteralgo}
The total number of variables in the cost function of the previous method is $2s_p s_q$, before any pre-processing.
Here we present the column-based algorithm where the number of variables (before pre-processing) is bounded by
$1+ s_p s_q + log_2(s_p) (s_p + s_q) $.   
Recall that here we are phrasing the factorization problem $M=pq$ as
	$$
	(\mathcal P_2):\,   
	min_{P_1, \dots, P_{sp}, Q_1, \dots, Q_{sq},Z_{12}, Z_{23}, Z_{24}, \ldots \in \mathbb Z_2}\,
	 \sum_{i} {H_{i}} ^2,
$$
where  $H_i$, for $1\leq i\leq s_p,$ is 
$$
H_i = \sum\limits_{j = 0}^{sq} {{Q_j}} {P_{i - j}} + \sum\limits_{j = 1}^i {{Z_{j,i}} - {m_i}}  - \sum\limits_{j = 1}^{ {L_i}} {{2^{j - i}}{Z_{i, i+j}}\begin{array}{*{20}{c}}
&{({Q_0} = {P_0} = {m_0} = 1, L_i = s_q +1 + i - {m_i})}.
\end{array}}
 $$
  The cost function is of degree 4 and, in order to use quantum annealing, it must be replaced with a positive
  quadratic polynomial with the same global minimum. The idea is to replace the quadratic terms ${Q_j} {P_{i - j}}$ inside the different $H_i$ with new binary variables $W_{i - j, j}$, and add the penalty  $({Q_j}P_{i - j} -W_{i - j, j})^+$ to the cost function (now written in terms of the variables $W_{i - j, j}$). To find $({Q_j} P_{i - j} -W_{i - j, j})^+$,  we run  Gr\"obner bases computation on the system 
  \begin{system}
  {Q_j} P_{i - j} -W, \\[2mm]
   {Q_j}^2-{Q_j}, \\[2mm]
   P_{i - j}^2-P_{i - j},  \\[2mm]
  {W_{i - j, j}}^2-{W_{i - j, j}}.
  \end{system}%
   Following the same steps as in the previous section, we get 
  $$({Q_j} P_{i - j} -W_{i - j, j})^+ = a({Q_j}W_{i - j, j}-W_{i - j, j})+
  b(P_{i - j} W_{i - j, j}-W_{i - j, j})+c(P_{i - j} {Q_j}-W_{i - j, j}), $$
   with  $a, b, c\in \mathbb R$ such that $-a-b-c>0, -b-c>0, -a-c>0, c>0$ (e.g., $c=1, a=b=-2$).  The new cost function is now
  $$ 
	\mathcal H = \sum_i  {H_i  (W)}^2 +  
	\sum _{ij} ({Q_j} P_{i - j} -W_{i - j, j})^+.
$$
We can obtain a better Hamiltonian by pre-processing
 the problem before applying the $W$ transformation. Indeed, let us first fix a positive integer ${\sf cutoff} \leq (s_p + s_q+1)$ and  let
	\mbox{$\mathcal B \subset  \mathbb{R}\left [P_1, \ldots,  P_{sp}, Q_1, \ldots, Q_{sq}, Z_{12}, Z_{23}, Z_{24} \ldots  \right ]$} be a Gr\"obner basis of the set of polynomials $$\{H_i\}_{i=1...{\sf cutoff}} \cup \{P_i (P_i-1), \, Q_i (Q_i-1),\,  Z_{ij} (Z_{ij}-1)\}_{i, j}.$$ 
In practice, the cutoff is determined by the size of the maximum subsystem of polynomials~$H_i$ on which one can run a Gr\"obner basis computation; it is defined
	by the hardware.  We also define a cutoff on the other tail of $\{H_i\}$, that is,
	we consider $\{H_i\}_{i={\sf 2ndcutoff}...(s_p+s_q+1)}$. Notice that here we are working on the original $H_i$ rather than the new $H_i(W)$.
	This is because we would like to perform the replacement ${Q_j} {P_{i - j}}\rightarrow  {W_{i - j, j}}$ after the pre-processing (some of the quadratic terms might be simplified by this pre-processing). 
	Precisely, what we pass to the quantum annealer is the quadratic positive polynomial 
\begin{equation}\label{finalH}
	\mathcal H = \sum \left({\sf NF}_{W_{i-j, j} -  {\sf LT} \left( {\sf NF}_{\mathcal B_c}  \left( Q_j P_{i - j}\right)\right)} \left ( {\sf NF} _{\mathcal B_c} (H_i) \right)\right)^2 +  
	\sum _{ij}  \left(W_{i-j, j} -  {\sf LT} \left( {\sf NF}_{\mathcal B_c}  \left( Q_j P_{i - j}\right)\right)\right)^+. 
\end{equation}
Here ${\sf LT}$ stands for the leading term with respect to the graded reverse lexicographic order. 
 The second summation is over all $i$ and $j$ such that   ${\sf LT} \left( {\sf NF}_\mathcal B  \left( Q_j P_{i - j}\right)\right)$
is still quadratic.  The outer normal form in the first summation refers to  the replacement  
${\sf LT} \left( {\sf NF}_\mathcal B  \left( Q_j P_{i - j}\right)\right)\rightarrow  {W_{i - j, j}}$, which is again performed only if
${\sf LT} \left( {\sf NF}_\mathcal B  \left( Q_j P_{i - j}\right)\right)$
is still quadratic.

~~\\ 
The columns of the following table present: a sample of bi-prime numbers and their prime factors,  the number of variables using each of a na\"ive polynomial-to-quadratic transformation {\it P2Q}, our novel polynomial-to-quadratic transformation {\it CustR}, and our window-based reduction {\it GB} after applying pre-processing.  The overall reduction percentage {\it R\%} and the embedding and solving status in the D-Wave 2X processor {\it Embed} are also shown. The adjacency matrix of the corresponding positive quadratic polynomial graph $\mathcal H$ and its embedded pattern inside the Chimera graph of the D-Wave 2X processor for one of the bi-primes are also depicted (see Figure 1).  Details pertaining to use of the hardware can be found in {\it Supplementary materials \ref{stats}}. 

\[\mathop {\boxed{\begin{array}{*{20}{c}}
  {\underline M }&{\underline {p \times q} }&{\underline {P2Q} }&{\underline {CustR} }&{\underline {GB} }&{\underline {R \textit{\%}} }&{\underline {Embed} } \\ 
  {150419}&{431 \times 349}&{116}&{86}&{73}&{37}&\surd  \\ 
  {151117}&{433 \times 349}&{117}&{88}&{72}&{38}&\surd  \\ 
  {174541}&{347 \times 503}&{117}&{86}&{72}&{38}&\surd  \\ 
  {200099}&{499 \times 401}&{115}&{89}&{75}&{35}&\surd  \\ 
  {223357}&{557 \times 401}&{125}&{96}&{80}&{36}& \times  
\end{array}}}\limits^{\begin{array}{*{20}{c}}
  {{\mathbf{Column}}}&{{\mathbf{Algorithm}}} 
\end{array}} \]

%

\begin{figure}[!htbp]\label{exples}
      \begin{minipage}{0.54\linewidth}
      \begin{center}
      \includegraphics[scale=0.495]{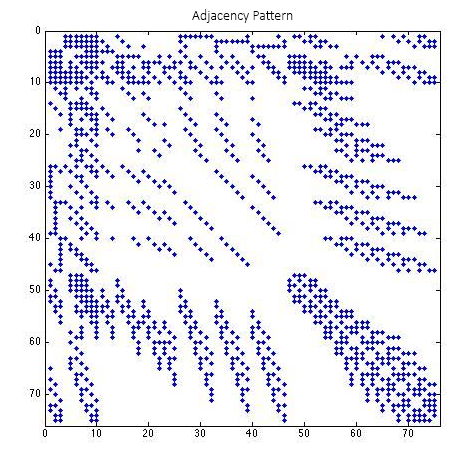}
      \end{center}
    \end{minipage}
    \begin{minipage}{0.4\linewidth}
      \includegraphics[scale=0.26]{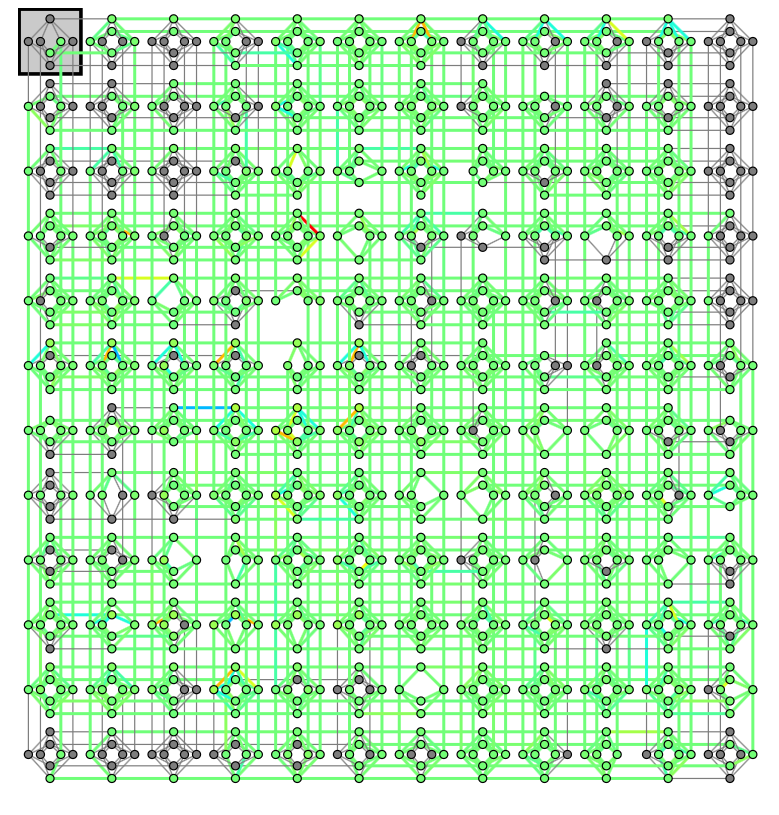} 
    \end{minipage}
    \caption{The column algorithm: the adjacency matrix pattern (left) and embedding into the the D-Wave 2X quantum processor (right) of the quadratic binary polynomial for {$M=$~\mbox{200 099}}.}
  \end{figure}

%

 \section{Discussion}
In this work, factorization is connected to quantum annealing  through binarization of the long multiplication.  The algorithm is autonomous in the sense that no a priori knowledge, or manual or ad hoc pre-processing, is involved. We have attained the largest bi-prime factored to date using a quantum processor, though more-subtle connections might exist. A future direction that this research can take is to connect  factorization (as an instance of the abelian hidden subgroup problem), through Galois correspondence,  to covering spaces and thus to covering graphs and potentially to quantum annealing. We believe that more-rewarding progress can be made through the investigation of such a connection.
 
\section{Methods}

\subsection{Column factoring procedure}
Here we discuss the two single-bit multiplication methods of the two primes $p$ and $q$. The first method  generates a Hamiltonian for each of the columns of the  long multiplication expansion, while the second method  generates a Hamiltonian for each of the multiplying cells in the long multiplication expansion.
The column factoring procedure of $p= 2^{sp} P_{sp} + 2^{sp-1} P_{sp-1} + ... + 2 P_1 + 1$ and $q = 2^{sq} Q_{sq} + 2^{sq-1} Q_{sq-1} ...+ 2 Q_1 + 1$  is depicted in the following table:
\\

\begin{LARGE}
\resizebox{0.97\linewidth}{0.15\linewidth}{
$
\begin{array}{*{20}{c}}
  {}&{}&{}&{}&{\begin{array}{*{20}{c}}
  {{P_{sp}}} \\ 
  {} 
\end{array}}&{\begin{array}{*{20}{c}}
   \cdots  \\ 
  {} 
\end{array}}&{\begin{array}{*{20}{c}}
  {{P_i}} \\ 
  {} 
\end{array}}&{\begin{array}{*{20}{c}}
   \cdots  \\ 
  {} 
\end{array}}&{\begin{array}{*{20}{c}}
  {{P_{sq}}} \\ 
  {{Q_{sq}}} 
\end{array}}&{\begin{array}{*{20}{c}}
   \cdots  \\ 
   \cdots  
\end{array}}&{\begin{array}{*{20}{c}}
  {{P_2}} \\ 
  {{Q_2}} 
\end{array}}&{\begin{array}{*{20}{c}}
  {{P_1}} \\ 
  {{Q_1}} 
\end{array}}&{\begin{array}{*{20}{c}}
  {{P_0} = 1} \\ 
  {{Q_0} = 1} 
\end{array}} \\ 
  {}&{\boxed{\begin{array}{*{20}{c}}
  {} \\ 
  {} \\ 
  {} \\ 
  {} \\ 
  {} \\ 
  {{Q_{sq}}{P_{sp}}} 
\end{array}}}&{\boxed{\begin{array}{*{20}{c}}
  {} \\ 
  {} \\ 
  {} \\ 
  {} \\ 
  {{Q_{sq - 1}}{P_{sp}}} \\ 
  {{Q_{sq}}{P_{sp - 1}}} 
\end{array}}}&{\begin{array}{*{20}{c}}
  {} \\ 
  {} \\ 
   {\mathinner{\mkern2mu\raise1pt\hbox{.}\mkern2mu
 \raise4pt\hbox{.}\mkern2mu\raise7pt\hbox{.}\mkern1mu}}  \\ 
   \cdots  \\ 
   \cdots  \\ 
   \cdots  
\end{array}}&{\boxed{\begin{array}{*{20}{c}}
  {{P_{sp}}} \\ 
  {{Q_1}{P_{sp - 1}}} \\ 
  {{Q_2}{P_{sp - 2}}} \\ 
   \vdots  \\ 
  {{Q_{sq - 1}}{P_{sp - sq + 1}}} \\ 
  {{Q_{sq}}{P_{sp - sq}}} 
\end{array}}}&{\begin{array}{*{20}{c}}
   \cdots  \\ 
   \cdots  \\ 
   \cdots  \\ 
   \cdots  \\ 
   \cdots  \\ 
   \cdots  
\end{array}}&{\boxed{\begin{array}{*{20}{c}}
  {{P_i}} \\ 
  {{Q_1}{P_{i - 1}}} \\ 
  {{Q_2}{P_{i - 2}}} \\ 
   \vdots  \\ 
  {{Q_{sq - 1}}{P_{i - sq + 1}}} \\ 
  {{Q_{sq}}{P_{i - sq}}} 
\end{array}}}&{\begin{array}{*{20}{c}}
   \cdots  \\ 
   \cdots  \\ 
   \cdots  \\ 
   \cdots  \\ 
   \cdots  \\ 
   \cdots  
\end{array}}&{\boxed{\begin{array}{*{20}{c}}
  {{P_{sq}}} \\ 
  {{Q_1}{P_{sq - 1}}} \\ 
  {{Q_2}{P_{sq - 2}}} \\ 
   \vdots  \\ 
  {{Q_{sq - 1}}{P_1}} \\ 
  {{Q_{sq}}} 
\end{array}}}&{\begin{array}{*{20}{c}}
   \cdots  \\ 
   \cdots  \\ 
   \cdots  \\ 
   \cdots  \\ 
   {\mathinner{\mkern2mu\raise1pt\hbox{.}\mkern2mu
 \raise4pt\hbox{.}\mkern2mu\raise7pt\hbox{.}\mkern1mu}}  \\ 
  {} 
\end{array}}&{\boxed{\begin{array}{*{20}{c}}
  {{P_2}} \\ 
  {{Q_1}{P_1}} \\ 
  {{Q_2}} \\ 
  {} \\ 
  {} \\ 
  {} 
\end{array}}}&{\boxed{\begin{array}{*{20}{c}}
  {{P_1}} \\ 
  {{Q_1}} \\ 
  {} \\ 
  {} \\ 
  {} \\ 
  {} 
\end{array}}}&{\boxed{\begin{array}{*{20}{c}}
  1 \\ 
  {} \\ 
  {} \\ 
  {} \\ 
  {} \\ 
  {} 
\end{array}}} \\ 
  {{m_{sp + sq + 1}}}&{{m_{sp + sq}}}&{{m_{sp + sq - 1}}}& \cdots &{{m_{sp}}}& \cdots &{{m_i}}& \cdots &{{m_{sq}}}& \cdots &{{m_2}}&{{m_1}}&{{m_0} = 1} 
\end{array}
$
}
\end{LARGE}
\\[3mm]
The equation for an arbitrary column ($i$) can be written as the sum of the column's  multiplication terms (above) plus all previously generated carry-on terms from lower significant columns ($j<i$). This sum is equal to the column's bi-prime term $m_i$ plus the carry-ons generated from higher significant columns. The polynomial equation for the $i$-th column is

\[\begin{array}{*{20}{c}}
  {\sum\limits_{j = 0}^{sq} {{Q_j}} {P_{i - j}} + \sum\limits_{j = 1}^i {{Z_{j,i}}}  = {m_i} + \sum\limits_{j = 1}^{{L_i}} {{2^{j - i}}} {Z_{i,i + j}}}&{\left( {{Q_0} = {P_0} = {m_0} = 1} \right).} 
\end{array}\]
The above equation is used as the main column procedure's equation $H_i$. The Hamiltonian generation and reduction is discussed in detail in \emph{Results \ref{betteralgo}}.   

\subsection{Cell bi-prime factoring procedure}
In the cell multiplication procedure, the ultimate goal is to break each of the column equations discussed above into multiple smaller equations so that each equation contains only one quadratic term. This not only simplifies the generation of quadratic Hamiltonians, but also generates Hamiltonians with more-uniform quadratic coefficients in comparison to the column procedure. The following table depicts the structure of the cell procedure:

{\color{white}  . }

 \begin{LARGE}
\resizebox{0.97\linewidth}{0.27\linewidth}{
$
\begin{array}{*{20}{c}}
  {}&{}&{}&{}&{\begin{array}{*{20}{c}}
  {{P_{sp}}} \\ 
  {} 
\end{array}}&{\begin{array}{*{20}{c}}
  {\begin{array}{*{20}{c}}
   \cdots &{{P_{i + j}}}& \cdots  
\end{array}} \\ 
  {} 
\end{array}}&{\begin{array}{*{20}{c}}
  {{P_{sq}}} \\ 
  {{Q_{sq}}} 
\end{array}}&{\begin{array}{*{20}{c}}
   \cdots  \\ 
   \cdots  
\end{array}}&{\begin{array}{*{20}{c}}
  {{P_3}} \\ 
  {{Q_3}} 
\end{array}}&{\begin{array}{*{20}{c}}
  {{P_2}} \\ 
  {{Q_2}} 
\end{array}}&{\begin{array}{*{20}{c}}
  {{P_1}} \\ 
  {{Q_1}} 
\end{array}}&{\begin{array}{*{20}{c}}
  {{P_0} = 1} \\ 
  {{Q_0} = 1} 
\end{array}} \\ 
  {}&{}&{}&{\boxed{\begin{array}{*{20}{c}}
  0 \\ 
  {{Q_1}{P_n}} \\ 
  {{Z_{1,n}}} 
\end{array}}}&{\boxed{\begin{array}{*{20}{c}}
  {{P_n}} \\ 
  {{Q_1}{P_{n - 1}}} \\ 
  {{Z_{1,n - 1}}} 
\end{array}}}& \vdots &{\boxed{\begin{array}{*{20}{c}}
  {{P_{sq}}} \\ 
  {{Q_1}{P_{sq}}} \\ 
  {{Z_{1,3}}} 
\end{array}}}& \cdots &{\boxed{\begin{array}{*{20}{c}}
  {{P_3}} \\ 
  {{Q_1}{P_2}} \\ 
  {{Z_{1,2}}} 
\end{array}}}&{\boxed{\begin{array}{*{20}{c}}
  {{P_2}} \\ 
  {{Q_1}{P_1}} \\ 
  {{Z_{1,1}}} 
\end{array}}}&{\boxed{\begin{array}{*{20}{c}}
  {{P_1}} \\ 
  {{Q_1}1} \\ 
  0 
\end{array}}}&{{m_0} = 1} \\ 
  {}&{}&{\boxed{\begin{array}{*{20}{c}}
  {{Z_{1,n + 1}}} \\ 
  {{Q_2}{P_n}} \\ 
  {{Z_{2,n}}} 
\end{array}}}&{\boxed{\begin{array}{*{20}{c}}
  {{S_{25}}} \\ 
  {{Q_2}{P_{n - 1}}} \\ 
  {{Z_{2,n - 1}}} 
\end{array}}}&{\boxed{\begin{array}{*{20}{c}}
  {{S_{24}}} \\ 
  {{Q_2}{P_4}} \\ 
  {{Z_{2,n - 2}}} 
\end{array}}}& \vdots &{\boxed{\begin{array}{*{20}{c}}
  {{S_{2,sq - 1}}} \\ 
  {{Q_2}{P_{sq - 1}}} \\ 
  {{Z_{2,sq - 1}}} 
\end{array}}}& \cdots &{\begin{array}{*{20}{c}}
  {\boxed{\begin{array}{*{20}{c}}
  {{S_{2,1}}} \\ 
  {{Q_2}{P_1}} \\ 
  {{Z_{2,1}}} 
\end{array}}} \\ 
  {} 
\end{array}}&{\begin{array}{*{20}{c}}
  {\boxed{\begin{array}{*{20}{c}}
  {{S_{20}}} \\ 
  {{Q_2}} \\ 
  0 
\end{array}}} \\ 
  {{m_2}} 
\end{array}}&{\begin{array}{*{20}{c}}
  {\begin{array}{*{20}{c}}
  {\begin{array}{*{20}{c}}
  {\begin{array}{*{20}{c}}
  {\begin{array}{*{20}{c}}
  {\begin{array}{*{20}{c}}
  {\begin{array}{*{20}{c}}
  {{m_1}} \\ 
  {} 
\end{array}} \\ 
  {} 
\end{array}} \\ 
  {} 
\end{array}} \\ 
  {} 
\end{array}} \\ 
  {} 
\end{array}} \\ 
  {} 
\end{array}} \\ 
  {} 
\end{array}}&{} \\ 
  {}& {\mathinner{\mkern2mu\raise1pt\hbox{.}\mkern2mu
 \raise4pt\hbox{.}\mkern2mu\raise7pt\hbox{.}\mkern1mu}} & \vdots & \vdots & \vdots &{\begin{array}{*{20}{c}}
  {\begin{array}{*{20}{c}}
  {} \\ 
  {} \\ 
  {{Z_{i,j + 1}}} 
\end{array}}&{\begin{array}{*{20}{c}}
  {} \\ 
  {\boxed{\begin{array}{*{20}{c}}
  {{S_{i,j}}} \\ 
  {{Q_i}{P_j}} \\ 
  {{Z_{i,j}}} 
\end{array}}} \\ 
  {{S_{i + 1,j - 1}}} 
\end{array}}&{} 
\end{array}}& \vdots & \vdots & {\mathinner{\mkern2mu\raise1pt\hbox{.}\mkern2mu
 \raise4pt\hbox{.}\mkern2mu\raise7pt\hbox{.}\mkern1mu}} &{}&{}&{} \\ 
  {}&{\boxed{\begin{array}{*{20}{c}}
  {{Z_{sq - 2,sp + 1}}} \\ 
  {{Q_{sq - 1}}{P_{sp}}} \\ 
  {{Z_{sq - 1,sp}}} 
\end{array}}}&{\boxed{\begin{array}{*{20}{c}}
  {{S_{sq - 1,sp - 1}}} \\ 
  {{Q_{sq - 1}}{P_{sp - 1}}} \\ 
  {{Z_{sq - 1,sp - 1}}} 
\end{array}}}&{\boxed{\begin{array}{*{20}{c}}
  {{S_{sq - 1,sp - 2}}} \\ 
  {{Q_{sq - 1}}{P_{sp - 2}}} \\ 
  {{Z_{sq - 1,sp - 2}}} 
\end{array}}}&{\boxed{\begin{array}{*{20}{c}}
  {{S_{sq - 1,sp - 3}}} \\ 
  {{Q_{sq - 1}}{P_{sp - 3}}} \\ 
  {{Z_{sq - 1,sp - 3}}} 
\end{array}}}& \vdots &{\boxed{\begin{array}{*{20}{c}}
  {{S_{sq - 1,1}}} \\ 
  {{Q_{sp - 1}}{P_1}} \\ 
  {{Z_{sq - 1,1}}} 
\end{array}}}&{\boxed{\begin{array}{*{20}{c}}
  {{S_{sq - 1,0}}} \\ 
  {{Q_{sq - 1}}} \\ 
  0 
\end{array}}}&{}&{}&{}&{} \\ 
  {\begin{array}{*{20}{c}}
  {\begin{array}{*{20}{c}}
  {} \\ 
  {} \\ 
  {} \\ 
  {{m_{sq + sp + 1}}} 
\end{array}}&{\boxed{\begin{array}{*{20}{c}}
  {{Z_{sq - 1,sp + 1}}} \\ 
  {{Q_{sq}}{P_{sp}}} \\ 
  {{Z_{sq,sp}}} 
\end{array}}} \\ 
  {}&{{m_{sq + sp}}} 
\end{array}}&{\begin{array}{*{20}{c}}
  {\boxed{\begin{array}{*{20}{c}}
  {{S_{sq,sp - 1}}} \\ 
  {{Q_{sq}}{P_{_{sp - 1}}}} \\ 
  {{Z_{sq,sp - 1}}} 
\end{array}}} \\ 
  {{m_{sq + sp - 1}}} 
\end{array}}&{\begin{array}{*{20}{c}}
  {\boxed{\begin{array}{*{20}{c}}
  {{S_{sq,sp - 2}}} \\ 
  {{Q_{sq}}{P_{sp - 2}}} \\ 
  {{Z_{sq,sp - 2}}} 
\end{array}}} \\ 
  {{m_{sq + sp - 2}}} 
\end{array}}&{\begin{array}{*{20}{c}}
  {\boxed{\begin{array}{*{20}{c}}
  {{S_{sq,sp - 3}}} \\ 
  {{Q_{sq}}{P_{sp - 3}}} \\ 
  {{Z_{sq,sp - 3}}} 
\end{array}}} \\ 
  {{m_{sq + sp - 3}}} 
\end{array}}&{\begin{array}{*{20}{c}}
  {\boxed{\begin{array}{*{20}{c}}
  {{S_{sq,sp - 4}}} \\ 
  {{Q_{sq}}{P_{sp - 4}}} \\ 
  {{Z_{sq,sp - 4}}} 
\end{array}}} \\ 
  {{m_{sq + sp - 4}}} 
\end{array}}&{\begin{array}{*{20}{c}}
  {} \\ 
   \vdots  \\ 
  {} \\ 
  {\begin{array}{*{20}{c}}
   \cdots &{{m_{i + j}}}& \cdots  
\end{array}} 
\end{array}}&{\begin{array}{*{20}{c}}
  {\boxed{\begin{array}{*{20}{c}}
  {{S_{sq0}}} \\ 
  {{Q_{sq}}} \\ 
  0 
\end{array}}} \\ 
  {{m_{sq}}} 
\end{array}}&{\begin{array}{*{20}{c}}
  {\begin{array}{*{20}{c}}
  {\begin{array}{*{20}{c}}
  {\begin{array}{*{20}{c}}
  {\begin{array}{*{20}{c}}
  {\begin{array}{*{20}{c}}
  {\begin{array}{*{20}{c}}
  {{m_{sq - 1}}} \\ 
  {} 
\end{array}} \\ 
  {} 
\end{array}} \\ 
  {} 
\end{array}} \\ 
  {} 
\end{array}} \\ 
  {} 
\end{array}} \\ 
  {} 
\end{array}} \\ 
  {} 
\end{array}}&{}&{}&{}&{} 
\end{array}
$
}
\end{LARGE}
\\[1mm]
Each cell contains one of the total $(s_p-1)(s_q-1)$ quadratic terms in the form of $Q_iP_j$. To chain a cell to its upper cell, one extra sum variable $S_{i,j}$ is added. Also, each carry-on variable $Z_{i,j}$ in a cell is the carry-on of the cell directly to its right, so each cell contains four variables. The sum of three terms $Q_iP_j$, $S_{i,j}$, and $Z_{i,j}$ is at most 3; thus, it generates an additional sum variable $S_{i+1,j-1}$ and one carry-on variable $Z_{i,j+1}$. Therefore, the equation for an arbitrary cell indexed $(i,j)$, shown in the centre of the above table, is

\[{S_{i,j}} + {Q_i}{P_j} + {Z_{i,j}} = {S_{i+1,j-1}} + 2{Z_{i,j+1}}.\]
As we can see, only six binary variables are involved in each cell equation and the equation contains one quadratic term, so it can be transformed into a positive Hamiltonian without adding slack variables. The Hamiltonian generation and reduction procedure is discussed in detail in \emph{Results \ref{cellalgo}}.

 
\subsection{Gr\"obner bases}
Good references for the following definitions are \cite{MR1363949, cox}.

~~\\
 {\it Normal forms.} A normal form is the remainder of Euclidean divisions in the ring of polynomials $\mathbb R[x_1, \ldots, x_n]$. Precisely, the normal form of a polynomial $f\in \mathbb R[x_1, \ldots, x_n]$, with respect to the set of polynomials
$\mathcal B\subset \mathbb R[x_1, \ldots, x_n]$ (usually a  Gr\"obner basis),  is the 
polynomial~${{\sf NF}(f)\in \mathbb R[x_1, \ldots, x_n]}$, which is the image of $f$ modulo $\mathcal B$. It is the remainder of the Euclidean of $f$ by all $g\in \mathcal B$.  

~~\\
 {\it Term orders.} A term order  on $\mathbb R[x_1, \ldots, x_n]$ is a total order $\prec$ on the set of all
monomials $x^a=x_1^{a_1}\ldots x_n^{a_n}$, which has the following properties: (1) 
 if $x^a\prec x^b$, then $x^{a+c}\prec x^{b+c}$ for all positive integers $a, b$, and $c$;  (2) $1\prec x^a$ for
all strictly positive integers $a$. An example of this is the pure lexicographic order $plex(x_1, \ldots, x_n)$. Monomials
are compared first by their degree in $x_1$, with ties broken by degree in $x_2$, etc. This order is usually used in eliminating variables.  Another example, is
the graded reverse lexicographic order $tdeg(x_1, \ldots, x_n)$. Monomials are compared first by their total degree, with ties broken by reverse lexicographic order, that is, by the smallest degree in $x_n, \, x_{n-1}$, etc. 
 
~~\\
{\it Gr\"obner bases.}
Given a term order $\prec$ on $\mathbb R[x_1, \ldots, x_n]$, then by the leading term (initial term) {\sf LT} of $f$ we mean the largest monomial in $f$ with respect to $\prec$. A (reduced) Gr\"obner basis to the ideal $\mathcal I$ with respect to the ordering~$\prec$ is a subset $\mathcal B$ of $\mathcal I$ such that: (1) the initial terms of elements of $\mathcal B$ generate the ideal ${\sf LT}(\mathcal I)$ of all initial terms of~$\mathcal I$; (2) for each $g\in \mathcal B$, the coefficient of the initial term of $g$ is 1; (3) the set ${\sf LT}(g)$ minimally generates ${\sf LT}(\mathcal I)$; and (4) no trailing term of any $g\in \mathcal B$ lies in ${\sf LT}(\mathcal I)$. Currently, Gr\"obner bases are computed using   sophisticated versions of the original Buchberger algorithm, for example, the F4 algorithm  by J. C. Faug\`ere.

\subsection{Factorization as an eigenvalue problem}
In this section, for completeness, we describe how the factorization problem can
be solved using eigenvalues and eigenvectors. This is an adaptation of the method presented in \cite{realalggeo} to factorization,
which is itself an adaption to real polynomial optimization of the method of solving polynomial equations using eigenvalues in~\cite{cox}.

~~\\
Let $\mathcal H$ be in $\mathbb R[x_1,\ldots, x_n]$ as in (\ref{finalH}), where we have used the notation
$x_i$ instead of the $Ps, Qs, Zs$, and $Ws$. Define
$$
	\mathcal H_\alpha := \mathcal H + \sum_i \alpha_i x_i(x_i-1),
$$
which is in  the larger ring $\mathbb R[x_1,\ldots, x_n,\alpha_1,\ldots, \alpha_n]$. We also define 
 the set of polynomials
$$
	\mathcal C = \left\{ \partial H_\alpha/\partial_{x_1},\ldots, \partial H_\alpha/\partial_{x_n},
	 \partial H_\alpha/\partial_{\alpha_1},\ldots, \partial H_\alpha/\partial_{\alpha_n}\right\}.  
$$
The variety $\mathcal V(\mathcal C)$ is the set of all binary critical points of $\mathcal H$.  Its
coordinates ring is the residue algebra $A:=\mathbb R[x_1,\ldots, x_n,\alpha_1,\ldots, \alpha_n]/\mathcal C$.
We need to compute a basis for $A$. This is done by first computing a Gr\"obner basis for $\mathcal C$ and then extracting the standard monomials (i.e., the monomials in 
$\mathbb R[x_1,\ldots, x_n,\alpha_1,\ldots, \alpha_n]$ that are not divisible by the leading term of any
element in the Gr\"obner basis).  In the simple example below, we do not need to compute any Gr\"obner basis since
 $\mathcal C$ is a Gr\"obner basis with respect to $plex (\alpha, x)$.   We define the linear map
\begin{eqnarray}\nonumber
m_{\mathcal H_\alpha }:& A& \rightarrow A\\\nonumber
							& g& \mapsto \mathcal H_\alpha  g	
\end{eqnarray}
Since the number of critical points is finite, the algebra 
$A$
 is always finite-dimensional by the Finiteness Theorem (\cite{cox}).
Now, the key points are:  
\begin{itemize}
	\item The value of $\mathcal H_\alpha$ (i.e., values of $\mathcal H$), on the set of critical points $\mathcal V(\mathcal C)$,
	are given by the  eigenvalues of the matrix $m_{\mathcal H_\alpha}$.
	\item Eigenvalues of $m_{x_i}$ and $m_{\alpha_i}$ give the coordinates of the points of  $\mathcal V(\mathcal C)$. 
	\item If $v$ is an eigenvector for $m_{\mathcal H_\alpha}$, then it is also an eigenvector for $m_{x_i}$ and $m_{\alpha_i}$ for $1\leq i\leq n$. 
\end{itemize}
We illustrate this in  an example. Consider $M=pq= 5\times 3$  and let 
$$
	\mathcal H = 2+7\,x_{{4}}+2\,x_{{3}}+2\,x_{{4}}x_{{3}}-2\,x_{{3}}x_{{2}}-x_{{1}}-4
\,x_{{4}}x_{{1}}-2\,x_{{3}}x_{{1}}+x_{{2}}x_{{1}}
$$
be the corresponding Hamiltonian as in (\ref{finalH}), where $x_{{1}}=p_{{2}},x_{{2}}=q_{{1}},x_{{3}}=w_{{2,1}}$,
and $x_4=z_{{2,3}}$.  A basis for the residue algebra $A$ is 
given by the set of the 16 monomials 
$$
\{ 1,x_{{4}},x_{{3}},x_{{4}}x_{{3}},x_{{2}},x_{{4}}x_{{2}},x_{{3}}x_{{2}
},x_{{3}}x_{{2}}x_{{4}},x_{{1}},x_{{4}}x_{{1}},x_{{3}}x_{{1}},x_{{1}}x
_{{3}}x_{{4}},x_{{2}}x_{{1}},x_{{4}}x_{{1}}x_{{2}},x_{{1}}x_{{3}}x_{{2
}},x_{{1}}x_{{3}}x_{{2}}x_{{4}}\}.
$$
The matrix $m_{\mathcal H_\alpha}$ is  
$$
m_{\mathcal H_\alpha} :=  \left[ \begin {array}{cccccccccccccccc} 2&7&2&2&0&0&-2&0&-1&-4&-2&0&1
&0&0&0\\ \noalign{\medskip}0&9&0&4&0&0&0&-2&0&-5&0&-2&0&1&0&0
\\ \noalign{\medskip}0&0&4&9&0&0&-2&0&0&0&-3&-4&0&0&1&0
\\ \noalign{\medskip}0&0&0&13&0&0&0&-2&0&0&0&-7&0&0&0&1
\\ \noalign{\medskip}0&0&0&0&2&7&0&2&0&0&0&0&0&-4&-2&0
\\ \noalign{\medskip}0&0&0&0&0&9&0&2&0&0&0&0&0&-4&0&-2
\\ \noalign{\medskip}0&0&0&0&0&0&2&9&0&0&0&0&0&0&-2&-4
\\ \noalign{\medskip}0&0&0&0&0&0&0&11&0&0&0&0&0&0&0&-6
\\ \noalign{\medskip}0&0&0&0&0&0&0&0&1&3&0&2&1&0&-2&0
\\ \noalign{\medskip}0&0&0&0&0&0&0&0&0&4&0&2&0&1&0&-2
\\ \noalign{\medskip}0&0&0&0&0&0&0&0&0&0&1&5&0&0&-1&0
\\ \noalign{\medskip}0&0&0&0&0&0&0&0&0&0&0&6&0&0&0&-1
\\ \noalign{\medskip}0&0&0&0&0&0&0&0&0&0&0&0&2&3&-2&2
\\ \noalign{\medskip}0&0&0&0&0&0&0&0&0&0&0&0&0&5&0&0
\\ \noalign{\medskip}0&0&0&0&0&0&0&0&0&0&0&0&0&0&0&5
\\ \noalign{\medskip}0&0&0&0&0&0&0&0&0&0&0&0&0&0&0&5\end {array}
 \right]
$$
We expect the matrix's smallest eigenvalue to be zero and, indeed, we get the following eigenvalues for $m_{\mathcal H_\alpha}$:
$$
\{0, 1, 2, 4, 5, 6, 9, 11, 13\}.
$$	 
This is also the set of values which $\mathcal H_\alpha$ takes on $\mathcal V(\mathcal C)$. The eigenvector $v$ which corresponds
to the eigenvalue 0 is the column vector
$$
	v := \left(1, 0, 1, 0, 1, 0, 1, 0, 1, 0, 1, 0, 1, 0, 1, 0 \right)^T.
$$
 This eigenvector is used to find the coordinates of $\hat x\in\mathcal V(\mathcal C)$ that cancel (minimize)~$\mathcal H_\alpha$. 
The coordinates of the global minimum $\hat x = (\hat x_1, \ldots, \hat x_n)$  are defined by $m_{x_i} v=\hat x_i v$, and this gives $x_1=x_2=x_3=1, \, x_4=0$, and $\alpha_1= 2\alpha_2 =\alpha_3=2, \, \alpha_4=5$. 

\newpage

\section{Supplementary materials} 

\subsection{Continuous optimization problems for the requirements (ii--iii)}
In \emph{Results 2.1}, 
we describe how a positive quadratic polynomial $H_{ij}^+$ can be extracted using Gr\"obner bases. Here we provide the details
of the calculation. 

~~\\ 
The second requirement (ii) is equivalent to each of the following linear polynomials being greater than zero:
\begin{normalsize}
\begin{eqnarray}\nonumber
 &&a_{{1}},-a_{{1}}+a_{{3}},-a_{{1}}-a_{{4}},-a_{{1}}+a_{{5}},-a
_{{1}}+a_{{6}},2\,a_{{1}}+a_{{3}},2\,a_{{1}}-a_{{4}},-a_{{2}}-a_{{1}},
-a_{{2}}+2\,a_{{1}},\\\nonumber
&& -2\,a_{{1}}+a_{{3}}+2\,a_{{5}},
-2\,a_{{1}}+a_{{3}}
+2\,a_{{6}},-2\,a_{{1}}-a_{{4}}+2\,a_{{5}}, -2\,a_{{1}}-a_{{4}}+2\,a_{{
6}},-a_{{1}}+a_{{5}}+a_{{6}},\\\nonumber
&&a_{{1}}+a_{{3}}-a_{{5}},a_{{1}}+a_{{3}}-a
_{{6}}, a_{{1}}-a_{{4}}-a_{{5}},a_{{1}}-a_{{4}}-a_{{6}},-a_{{2}}-2\,a_{
{1}}+2\,a_{{5}},-a_{{2}}-2\,a_{{1}}+2\,\\\nonumber
&&a_{{6}},-a_{{2}}+a_{{1}}-a_{{5}
},-a_{{2}}+a_{{1}}-a_{{6}}, -2\,a_{{1}}+a_{{3}}+2\,a_{{5}}+2\,a_{{6}},-
2\,a_{{1}}-a_{{4}}+2\,a_{{5}}+2\,a_{{6}},\\\nonumber
&&a_{{1}}+a_{{3}}-a_{{5}}-a_{{6
}},a_{{1}}-a_{{4}}-a_{{5}}-a_{{6}}, -a_{{2}}-2\,a_{{1}}+a_{{3}}-a_{{4}}
,-a_{{2}}-2\,a_{{1}}+2\,a_{{5}}+2\,a_{{6}},\\\nonumber
&&-a_{{2}}+a_{{1}}-a_{{5}}-a_
{{6}},-a_{{2}}+3\,a_{{1}}+a_{{3}}-a_{{4}}, -a_{{2}}-3\,a_{{1}}+a_{{3}}-
a_{{4}}+3\,a_{{5}},\\\nonumber
&& -a_{{2}}-3\,a_{{1}}+a_{{3}}-a_{{4}}+3\,a_{{6}},-a_{
{2}}+2\,a_{{1}}+a_{{3}}-a_{{4}}-2\,a_{{5}},-a_{{2}}+2\,a_{{1}}+a_{{3}}
-a_{{4}}-2\,a_{{6}},\\\nonumber
&& -a_{{2}}-3\,a_{{1}}+a_{{3}}-a_{{4}}+3\,a_{{5}}+3\,
a_{{6}},-a_{{2}}+2\,a_{{1}}+a_{{3}}-a_{{4}}-2\,a_{{5}}-2\,a_{{6}}
\end{eqnarray}
\end{normalsize}
For the third requirement (iii), the first choice for the objective function $f:\mathbb R^5\rightarrow \mathbb R$ is 
\begin{eqnarray}\nonumber
f(a_1, \ldots, a_5) &=&
 \left(  \left( -a_{{1}}+a_{{5}}+a_{{6}} \right) ^{2}-1 \right) ^{2}+
 \left(  \left( -2\,a_{{1}}+a_{{3}}+2\,a_{{5}}+2\,a_{{6}} \right) ^{2}
-1 \right) ^{2}\\\nonumber
&+& \left(  \left( a_{{1}}-a_{{2}}-a_{{5}}-a_{{6}}
 \right) ^{2}-1 \right) ^{2}+ \left(  \left( a_{{1}}-a_{{4}}-a_{{5}}-a
_{{6}} \right) ^{2}-1 \right) ^{2}\\\nonumber
&+& 2\, \left( {a_{{2}}}^{2}-1 \right) 
^{2}+ \left( {a_{{1}}}^{2}-1 \right) ^{2}+2\, \left( {a_{{3}}}^{2}-1
 \right) ^{2}+2\, \left( {a_{{4}}}^{2}-1 \right) ^{2}\\\nonumber
 &+& 2\, \left( {a_{{
5}}}^{2}-1 \right) ^{2}+2\, \left( {a_{{6}}}^{2}-1 \right) ^{2}
+ \left( 4\,{a_{{5}}}^{2}-1 \right) ^{2}+ \left( 4\,{a_{{6}}}^{2}-1\right) ^{2}
\end{eqnarray}
The solution is $a_{{1}}= 0.214, \, a_{{2}}=- 1.082,  a_{{3}}=
 0.514,a_{{4}}=- 1.082,\, a_{{5}}=
 0.314,$ and $a_{{6}}= 0.314. $

~~\\ 
The second choice for $f$ is
\begin{eqnarray}\nonumber
 f(a_1, \ldots, a_5) &=& 
 \left(  \left( -a_{{1}}+a_{{5}}+a_{{6}} \right) ^{2}-a_{{2}} \right) 
^{2}+ \left(  \left( -2\,a_{{1}}+a_{{3}}+2\,a_{{5}}+2\,a_{{6}}
 \right) ^{2}-a_{{2}} \right) ^{2}\\\nonumber
 &+&  \left(  \left( a_{{1}}-a_{{2}}-a_{
{5}}-a_{{6}} \right) ^{2}-a_{{2}} \right) ^{2}+ \left(  \left( a_{{1}}
-a_{{4}}-a_{{5}}-a_{{6}} \right) ^{2}-a_{{2}} \right) ^{2}\\\nonumber
&+& 2\, \left( 
{a_{{2}}}^{2}-a_{{2}} \right) ^{2}+ \left( {a_{{1}}}^{2}-a_{{2}}
 \right) ^{2}+2\, \left( {a_{{3}}}^{2}-a_{{2}} \right) ^{2}+2\,
 \left( {a_{{4}}}^{2}-a_{{2}} \right) ^{2}\\
 &+& 2\, \left( {a_{{5}}}^{2}-a_
{{2}} \right) ^{2}+2\, \left( {a_{{6}}}^{2}-a_{{2}} \right) ^{2}+
 \left( 4\,{a_{{5}}}^{2}-a_{{2}} \right) ^{2}+ \left( 4\,{a_{{6}}}^{2}
-a_{{2}} \right) ^{2}
\end{eqnarray}
The solution is $a_{{1}}= 1.0,a_{{2}}=- 4.0,a_{{3}}= 4.0,a_{{4}}=- 4.0,a_{{5}}= 2.0,$ and $a_
{{6}}= 2.0$ (identical to the solution given in \cite{DBLP:journals/qic/SchallerS10}). 
 
\subsection{Basic description of the quantum annealing processor}
Here we introduce the quantum annealing concept that ultimately solves a general Ising (quadratic unconstrained binary optimization, or "QUBO") problem, then  talk about the important topic of embedding a QUBO problem into the specific quantum annealer
(the D-Wave 2X processor).

~~\\
Quantum annealing (QA), along with the D-Wave processors, have been the focus of much research. We refer the interested reader to 
\cite{natueDwave, Calude:2015:GCA:2744447.2744459, BoixoNature, BoixoNature2, PhysRevX.4.021041}.  
QA is a paradigm designed to find the ground state of systems of interacting spins represented by a time-evolving Hamiltonian:
$$
	\mathcal S(s) =\mathcal E(s)\mathcal H_P - \frac{1}{2} \sum_i \Delta(s)\sigma_i^x,
$$
$$
	\mathcal H_P = -\sum_i h_i \sigma_i^x + \sum_{i<j}J_{ij}\sigma_i^z\sigma_j^z.
$$
The parameters $h_i$ and $J_{ij}$ encode the particular QUBO problem $P$ into its {Ising} formulation. QA is performed by first setting $\Delta \gg \mathcal E,$ which results in a ground state into which the spins can be easily initialized. Then $\Delta$ is {\it slowly} reduced and $\mathcal E$ is increased until $\mathcal E\gg \Delta$. At this point the system is dominated by $\mathcal H_P,$ which encodes the optimization problem. Thus, the ground state represents the solution to the optimization problem.  

~~\\
An embedding is the mapping of the nodes of an input graph to the nodes of the destination graph. The graph representing the problem's QUBO matrix needs to be embedded into the actual physical qubits on the processor in order for it to solve the QUBO problem. The specific existing connectivity pattern of qubits in the D-Wave chip is called the Chimera graph. Embedding an input graph (a QUBO problem graph) into the hardware graph (the Chimera graph) 
is in general NP-hard~{(\cite{Choi}).} 

 ~~\\
Figure 1--right  shows an embedding of the (column algorithm) QUBO  corresponding to the bi-prime $M=$~\mbox{200 099} 
into the Chimera graph of the D-Wave 2X chip consisting of a 12 by 12 lattice of 4 by 4 bipartite blocks. The Chimera graph is structured so that the vertical and horizontal couplers in its lattice are connected only to either side of each bipartite block.  Each node in this graph represents one qubit and each edge represents a coupling between two qubits. 
Adjacent nodes in the Chimera graph can be grouped together to form new effective (i.e., logical) nodes, creating nodes of a higher degree. Such a grouping is performed on the processor by setting the coupler between two qubits to a large negative value, forcing two Ising spins to align such that the two qubits end up with the same values. These effective qubits are expected to behave identically and remain in the same binary state at the time of measurement. The act of grouping adjacent qubits (hence forming new effective qubits) is called chain creation or chain identification. 
 
  ~~\\
  An embedding strategy consists of two tasks: mapping and identification.  Mapping is the assignment of the nodes of the input graph to the single or effective nodes of the destination graph. Solving such problems optimally is in general NP-hard, but one can devise various approximations and enhancement strategies to overcome these difficulties, for example, using statistical search methods like simulated annealing, structure-based methods, or a combination of both. For a better understanding of current embedding approaches, we refer the reader to
   \cite{Choi, roy, king0, king1}. In Figure 1--right, the blue lines indicate the identified couplers, the yellow lines indicate the problem couplers (i.e., the edges of the problem graph), and the grey lines indicate empty couplers.

\subsection{Embedding and solving details}\label{stats}
We have used one of the D-Wave 2X processors, DW2X\_SYS4, as our quantum annealing solver. This processor operates at a temperature range of $26(\pm 5)$ millikelvin (mK) and has 1100  qubits with a 95.5-qubit yield. To utilize the processor, we used D-Wave's SAPI software development kit (version 2.2.1). To embed the problem graph into the hardware graph we used the sapiFindEmbedding and sapiEmbedProblem modules, and to solve the problems we used the sapiSolveIsing and sapiUnembedAnswer modules. For all problems we opted for the maximum number of reads available (\mbox{10 000}) in order to increase the fraction of ground state samples. The following table shows some statistics of the embedding and solving stages for several of the highest numbers that we were able to successfully embed and solve.

\[\mathop {\boxed{\begin{array}{*{20}{c}}
  {\underline M }&{\underline n }&{\underline {emTry} }&{\underline {idC} }&{\underline {prC} }&{\underline {\# qubits} }&{\underline {jRatio} }&{\underline {rTime} } \\ 
  {31861}&{95}&{33}&{848}&{721}&{815}&{10}&{3.52} \\ 
  {34889}&{95}&{27}&{803}&{740}&{833}&{10}&{3.52} \\ 
  {150419}&{73}&1&{941}&{830}&{902}&{64}&{3.52} \\ 
  {151117}&{72}&7&{1001}&{846}&{918}&{64}&{3.52} \\ 
  {174541}&{72}&3&{1004}&{897}&{966}&{64}&{3.52} \\ 
  {200099}&{75}&5&{884}&{824}&{897}&{64}&{3.52} 
\end{array}}}\limits^{Embedding \,\&\, Solving \, Statistics} \]
 In the above table, $M$ stands for the bi-prime, $n$ is the number of variables in the QUBO problem, {\it emTry} is the number of block trials of  the sapiFindEmbedding routine,   {\it idC} is the total number of identified couplers, {\it prC} is the total number of problem couplers,
   {\it \#qubits} is the total number of (physical) qubits, {\it jRatio} is the ratio $\frac{\mathrm{max}(\{|J_{ij}|\})}{\mathrm{min}(\{|J_{ij}|\})}$, and 
   {\it rTime} is the chip run time in seconds.

\section{Acknowledgements} 
 {We would like to thank Pooya Ronagh for constructive discussions and helpful comments on the paper. We also thank Marko Bucyk for proofreading the manuscript.}

\bibliography{c}
\end{document}

%% file: macro.tex


\newcommand{\calA}{{\cal A}}
\newcommand{\calB}{{\cal B}}
\newcommand{\calF}{{\cal F}}
\newcommand{\calG}{{\cal G}}
\newcommand{\calR}{{\cal R}}
\newcommand{\calZ}{{\cal Z}}
\def\D{\mathcal{D}}
\def\L{\mathcal{L}}
\def\S{\mathcal{S}}
\def\I{\mathcal{I}}
\def\V{\mathcal{V}}
\def\E{\mathcal{E}}
\def\M{\mathcal{M}}

\def\A{\mathscr{A}}
\def\F{\mathscr{F}}
\def\G{\mathscr{G}}

\newcommand{\N}{{\mathbb N}}
\newcommand{\Z}{{\mathbb Z}}
\newcommand{\Q}{{\mathbb Q}}
\newcommand{\R}{{\mathbb R}}
\newcommand{\CC}{{\mathbb C}}

\newcommand{\K}{{\mathbb K}}
\newcommand{\kk}{{\mathrm k}}

\def\J{\mathrm{J}}
\def\catC{{\bf \mathrm{C}}}
\def\x{\mathrm{x}}
\def\a{\mathrm{a}}
\def\d{\mathrm{d}}
\def\Bi{\mathrm{Bi}}
\def\op{\mathrm{op}}
\def\res{\mathrm{res}}
\def\span{\mathrm{span}}

\newcommand{\C}{{\bf C}}
\newcommand{\Objects}{{\bf Objects}}
\newcommand{\Arrows}{{\bf Arrows}}
\newcommand{\Sets}{{\bf Sets}}

\def\2F1{\mbox{ $_2${F}$_1$}}
\def\1F1{\mbox{ $_1${F}$_1$}}
\def\1F2{\mbox{ $_1${F}$_2$}}
\def\0F1{\mbox{ $_0${F}$_1$}}

\def\GL{\mathrm{GL}}
\def\det{\mathrm{det}}
\def\SL{\mathrm{SL}}
\def\PSL{\mathrm{PSL}}
\def\PGL{\mathrm{PGL}}
\def\O{\mathrm{O}}

\def\gl{\mathfrak{gl}}
\def\g{\mathfrak{g}}
\def\h{\mathfrak{h}}
\def\frakM{\mathfrak{M}}

\newcommand{\Frac}[2]{\displaystyle \frac{#1}{#2}}
\newcommand{\Sum}[2]{\displaystyle{\sum_{#1}^{#2}}}
\newcommand{\Prod}[2]{\displaystyle{\prod_{#1}^{#2}}}
\newcommand{\Int}[2]{\displaystyle{\int_{#1}^{#2}}}
\newcommand{\Lim}[1]{\displaystyle{\lim_{#1}\ }}

\newenvironment{menumerate}{%
    \renewcommand{\theenumi}{\roman{enumi}}%
    \renewcommand{\labelenumi}{\rm(\theenumi)}%
    \begin{enumerate}} {\end{enumerate}}

\newenvironment{system}[1][]%
	{\begin{eqnarray} #1 \left\{ \begin{array}{lll}}%
	{\end{array} \right. \end{eqnarray}}

\newenvironment{meqnarray}%
	{\begin{eqnarray}  \begin{array}{rcl}}%
	{\end{array}  \end{eqnarray}}

\newenvironment{marray}%
	{\\ \begin{tabular}{ll}}
	{\end{tabular}\\}

\newenvironment{program}[1]%
	{\begin{center} \hrulefill \quad {\sf #1} \quad \hrulefill \\[8pt]
		\begin{minipage}{0.90\linewidth}}
	{\end{minipage} \end{center} \hrule \vspace{2pt} \hrule}

\newcommand{\entrylabel}[1]{\mbox{\textsf{#1:}}\hfil}
\newenvironment{entry}
   {\begin{list}{}%
   	{\renewcommand{\makelabel}{\entrylabel}%
   	  \setlength{\labelwidth}{40pt}%
   	  \setlength{\leftmargin}{\labelwidth + \labelsep}%
   	}%
   }%
   {\end{list}}

\newenvironment{remark}{\par \noindent {\bf Remark. }}
			{\hfill $\blacksquare$ \par}
\newenvironment{example}{\par \noindent {\bf Example. }}
			{\hfill $\blacksquare$ \par}

\newenvironment{Pmatrix}
        {$ \left( \!\! \begin{array}{rr} }
        {\end{array} \!\! \right) $}

\newcommand{\fleche}[1]{\stackrel{#1}\longrightarrow}
\def\ssi{si et seulement si\ }
\newcommand{\tab}{\hspace*{\fill}}
\newcommand{\bs}{{\backslash}}
\newcommand{\eps}{{\varepsilon}}
\newcommand{\into}{{\;\rightarrow\;}}
\newcommand{\PD}[2]{\frac{\partial #1}{\partial #2}}
\def\Hat{\widehat}
\def\Bar{\overline}
\def\vect{\vec}
\def\fbar{{\bar f}}
\def\xbar{{\bar \x}}
\newcommand{\afaire}[1]{$$\vdots$$ \begin{center} {\sc #1} \end{center} $$\vdots$$ }
\newcommand{\pref}[1]{(\ref{#1})}

\def\Maple{{\sc Maple}}
\def\RG{{\sc Rosenfeld-Gr\"obner}}



\newcommand{\algf}{\sffamily}
\newcommand{\BEGIN}{{\algf begin}}
\newcommand{\END}{{\algf end}}
\newcommand{\IF}{{\algf if}}
\newcommand{\THEN}{{\algf then}}
\newcommand{\ELSE}{{\algf else}}
\newcommand{\ELIF}{{\algf elif}}
\newcommand{\FI}{{\algf fi}}
\newcommand{\WHILE}{{\algf while}}
\newcommand{\FOR}{{\algf for}}
\newcommand{\DO}{{\algf do}}
\newcommand{\OD}{{\algf od}}
\newcommand{\RETURN}{{\algf return}}
\newcommand{\PROCEDURE}{{\algf procedure}}
\newcommand{\FUNCTION}{{\algf function}}
\newcommand{\INDENTER}{{\algf si} \=\+\kill}

\newcommand{\target}{\mathop{\mathrm{t}}}
\newcommand{\source}{\mathop{\mathrm{s}}}
\newcommand{\trdeg}{\mathop{\mathrm{tr~deg}}}
\newcommand{\jet}[2]{\jmath_{#1}^{#2}}
\newcommand{\rank}{\operatorname{rank}}
\newcommand{\sign}{\operatorname{sign}}
\newcommand{\ord}{\operatorname{ord}}
\newcommand{\aut}{\operatorname{aut}}
\newcommand{\Hom}{\operatorname{Hom}}
\newcommand{\myhom}{\operatorname{hom}}
\newcommand{\codim}{\operatorname{codim}}
\newcommand{\coker}{\operatorname{coker}}
\newcommand{\rp}{\operatorname{rp}}
\newcommand{\leader}{\operatorname{ld}}
\newcommand{\card}{\operatorname{card}}
\newcommand{\Fr}{\operatorname{Frac}}
\newcommand{\RF}{\operatorname{\mathsf{reduced\_form}}}
\newcommand{\rang}{\operatorname{rang}}

\def \Id{\mathrm{Id}}

\def \diff{\mathrm{Diff}^{\mathrm{loc}} }
\def \diffg{\mathrm{Diff} }
\def \Esc{\mathrm{Esc}}

\newcommand{\initial}{\mathop{\mathsf{init}}}
\newcommand{\separant}{\mathop{\mathsf{sep}}}
\newcommand{\quo}{\mathop{\mathsf{quo}}}
\newcommand{\pquo}{\mathop{\mathsf{pquo}}}
\newcommand{\lcoeff}{\mathop{\mathsf{lcoeff}}}
\newcommand{\mvar}{\mathop{\mathsf{mvar}}}

\newcommand{\prem}{\mathop{\mathsf{prem}}}
\newcommand{\remp}{\mathrel{\mathsf{partial\_rem}}}
\newcommand{\remf}{\mathrel{\mathsf{full\_rem}}}
\renewcommand{\gcd}{\mathop{\mathrm{gcd}}}
\newcommand{\pairs}{\mathop{\mathrm{pairs}}}
\newcommand{\dd}{\mathrm{d}}
\newcommand{\ideal}[1]{(#1)}
\newcommand{\cont}{\mathop{\mathrm{cont}}}
\newcommand{\pp}{\mathop{\mathrm{pp}}}
\newcommand{\pgcd}{\mathop{\mathrm{pgcd}}}
\newcommand{\ppmc}{\mathop{\mathrm{ppcm}}}
\newcommand{\init}{\mathop{\mathrm{initial}}}

%% file: 1QBit_Prime_Factorization_Research_Paper.bbl
\newcommand{\etalchar}[1]{$^{#1}$}
\def\cprime{$'$}
\providecommand{\bysame}{\leavevmode\hbox to3em{\hrulefill}\thinspace}
\providecommand{\MR}{\relax\ifhmode\unskip\space\fi MR }
\providecommand{\MRhref}[2]{%
  \href{http://www.ams.org/mathscinet-getitem?mr=#1}{#2}
}
\providecommand{\href}[2]{#2}
\begin{thebibliography}{MNM{\etalchar{+}}16}

\bibitem[BAS{\etalchar{+}}13]{BoixoNature}
Sergio Boixo, Tameem Albash, Federico~M. Spedalieri, Nicholas Chancellor, and
  Daniel~A. Lidar, \emph{Experimental signature of programmable quantum
  annealing}, Nat Commun \textbf{4} (2013).

\bibitem[BCI{\etalchar{+}}14]{roy}
Zhengbing Bian, Fabian Chudak, Robert Israel, Brad Lackey, William~G Macready,
  and Aidan Roy, \emph{Discrete optimization using quantum annealing on sparse
  ising models}, Frontiers in Physics \textbf{2} (2014), no.~56.

\bibitem[BRI{\etalchar{+}}14]{BoixoNature2}
Sergio Boixo, Troels~F. Ronnow, Sergei~V. Isakov, Zhihui Wang, David Wecker,
  Daniel~A. Lidar, John~M. Martinis, and Matthias Troyer, \emph{Evidence for
  quantum annealing with more than one hundred qubits}, Nat Phys \textbf{10}
  (2014), no.~3, 218--224.

\bibitem[Bur02]{microsoft}
C.J.C. Burges, \emph{Factoring as optimization}, Tech. Report MSR-TR-2002-83,
  Microsoft Research, January 2002.

\bibitem[CCD15]{Calude:2015:GCA:2744447.2744459}
Cristian~S. Calude, Elena Calude, and Michael~J. Dinneen, \emph{Guest column:
  Adiabatic quantum computing challenges}, SIGACT News \textbf{46} (2015),
  no.~1, 40--61.

\bibitem[Cho08]{Choi}
Vicky Choi, \emph{Minor-embedding in adiabatic quantum computation: I. the
  parameter setting problem}, Quantum Information Processing \textbf{7} (2008),
  no.~5, 193--209.

\bibitem[CLO98]{cox}
David~A. Cox, John~B. Little, and Donal O'Shea, \emph{Using algebraic
  geometry}, Graduate texts in mathematics, Springer, New York, 1998.

\bibitem[DBI{\etalchar{+}}15]{citeulike:13881878}
Vasil~S. Denchev, Sergio Boixo, Sergei~V. Isakov, Nan Ding, Ryan Babbush, Vadim
  Smelyanskiy, John Martinis, and Hartmut Neven, Preprint arXiv:1512.02206
  (2015).

\bibitem[JAG{\etalchar{+}}11]{natueDwave}
M.~W. Johnson, M.~H.~S. Amin, S.~Gildert, T.~Lanting, F.~Hamze, N.~Dickson,
  R.~Harris, A.~J. Berkley, J.~Johansson, P.~Bunyk, E.~M. Chapple, C.~Enderud,
  J.~P. Hilton, K.~Karimi, E.~Ladizinsky, N.~Ladizinsky, T.~Oh, I.~Perminov,
  C.~Rich, M.~C. Thom, E.~Tolkacheva, C.~J.~S. Truncik, S.~Uchaikin, J.~Wang,
  B.~Wilson, and G.~Rose, \emph{Quantum annealing with manufactured spins},
  Nature \textbf{473} (2011), no.~7346, 194--198.

\bibitem[JWA14]{king0}
Cai Jun, G.~Macready William, and Roy Aidan, \emph{A practical heuristic for
  finding graph minors}, Preprint arXiv:1406.2741 (2014).

\bibitem[Kit95]{hsp}
A.Yu. Kitaev, \emph{Quantum measurements and the abelian stabilizer problem},
  quant-ph/9511026 (1995).

\bibitem[LPS{\etalchar{+}}14]{PhysRevX.4.021041}
T.~Lanting, A.~J. Przybysz, A.~Yu. Smirnov, F.~M. Spedalieri, M.~H. Amin, A.~J.
  Berkley, R.~Harris, F.~Altomare, S.~Boixo, P.~Bunyk, N.~Dickson, C.~Enderud,
  J.~P. Hilton, E.~Hoskinson, M.~W. Johnson, E.~Ladizinsky, N.~Ladizinsky,
  R.~Neufeld, T.~Oh, I.~Perminov, C.~Rich, M.~C. Thom, E.~Tolkacheva,
  S.~Uchaikin, A.~B. Wilson, and G.~Rose, \emph{Entanglement in a quantum
  annealing processor}, Phys. Rev. X \textbf{4} (2014), 021041.

\bibitem[MNM{\etalchar{+}}16]{Monz1068}
Thomas Monz, Daniel Nigg, Esteban~A. Martinez, Matthias~F. Brandl, Philipp
  Schindler, Richard Rines, Shannon~X. Wang, Isaac~L. Chuang, and Rainer Blatt,
  \emph{Realization of a scalable shor algorithm}, Science \textbf{351} (2016),
  no.~6277, 1068--1070.

\bibitem[PS01]{realalggeo}
Pablo~A. Parrilo and Bernd Sturmfels, \emph{Minimizing polynomial functions},
  DIMACS Series in Discrete Mathematics and Theoretical Computer Science
  (2001).

\bibitem[Rau13]{PhysRevA.88.022322}
Robert Raussendorf, \emph{Contextuality in measurement-based quantum
  computation}, Phys. Rev. A \textbf{88} (2013), 022322.

\bibitem[Sho97]{MR1471990}
Peter~W. Shor, \emph{Polynomial-time algorithms for prime factorization and
  discrete logarithms on a quantum computer}, SIAM J. Comput. \textbf{26}
  (1997), no.~5, 1484--1509. \MR{1471990}

\bibitem[SS10]{DBLP:journals/qic/SchallerS10}
Gernot Schaller and Ralf Schutzhold, \emph{The role of symmetries in adiabatic
  quantum algorithms}, Quantum Information {\&} Computation \textbf{10} (2010),
  no.~1{\&}2, 109--140.

\bibitem[SSV13]{smolin}
John~A. Smolin, Graeme Smith, and Alexander Vargo, \emph{Oversimplifying
  quantum factoring}, Nature \textbf{499} (2013), no.~7457, 163--165.

\bibitem[Stu96]{MR1363949}
Bernd Sturmfels, \emph{Gr\"obner bases and convex polytopes}, University
  Lecture Series, vol.~8, American Mathematical Society, Providence, RI, 1996.
  \MR{1363949}

\bibitem[TAA15]{king1}
Boothby Tomas, D.~King Andrew, and Roy Aidan, \emph{Fast clique minor
  generation in chimera qubit connectivity graphs}, Preprint arXiv:1507.04774
  (2015).

\bibitem[XZL{\etalchar{+}}12]{crapyPRL}
Nanyang Xu, Jing Zhu, Dawei Lu, Xianyi Zhou, Xinhua Peng, and Jiangfeng Du,
  \emph{Quantum factorization of 143 on a dipolar-coupling nuclear magnetic
  resonance system}, Phys. Rev. Lett. \textbf{108} (2012), 130501.

\end{thebibliography}
